

\font\twelvebf=cmbx12
\font\ninerm=cmr9
\nopagenumbers
\magnification =\magstep 1
\overfullrule=0pt
\baselineskip=18pt
\line{\hfil CCNY-HEP 2/95}
\line{\hfil March 1995}
\vskip .8in
\centerline{\twelvebf A Self-consistent Inclusion of Magnetic Screening }
\centerline{\twelvebf for the Quark-Gluon Plasma}
\vskip .5in
\centerline{\ninerm G. ALEXANIAN and V.P. NAIR \footnote {*}{E-mail:
garnik@scisun.sci.ccny.cuny.edu, vpn@ajanta.sci.ccny.cuny.edu}}
\vskip .1in
\centerline{ Physics Department}
\centerline{City College of the City University of New York}
\centerline{New York, New York 10031.}
\vskip 1in
\baselineskip=16pt
\centerline{\bf Abstract}
\vskip .1in
We discuss how magnetic screening can be systematically
included in a self-consistent way for Chromodynamics at high temperatures.
The resulting gap equation, which gives the
summation of self-energy insertions, is calculated to one-loop order and
leads to a nonzero value for
the magnetic mass.
\vfill\eject
\input epsf.tex
\footline={\hss\tenrm\folio\hss}
\def\af{\alpha}
\def\rtm{\sqrt{k^2(\alpha-\alpha^2)+M^2}}
\def\rtn{\sqrt{k^2(\alpha-\alpha^2)+M^2\alpha}}
\def\dlmb{\partial^2}
\def \12 {{\textstyle {1\over 2}}}
\def \vf {\varphi}
The generating functional for hard thermal loops in Quantum
Chromodynamics (QCD) is closely related to the eikonal for a
Chern-Simons theory [1]. The importance and various properties of hard
thermal loops have been the subject of many recent investigations [2-5].
The generalization of the Chern-Simons eikonal to a moving plasma
suggests that there is another closely related gauge-invariant mass
term which gives screening of static magnetic interactions, in
other words, a magnetic mass term [6]. Although nonlocal,
the equations of motion for
this term can be written in a local way by using auxiliary fields and without
introducing additional degrees of freedom. Because of this last property,
the ultraviolet behaviour of the theory is unchanged. On this basis,
it was suggested
that this mass term could be used as a gauge-invariant infrared
cutoff for loop calculations in QCD at high temperatures. In
this paper we discuss how this can be carried out systematically.
Some recent related papers are listed as ref. [7].

A consistent implementation of a gauge-invariant infrared cutoff
will lead to a gap equation for the value of the magnetic mass.
We obtain this equation to one-loop order; as usual, this means a
self-consistent summation of one loop self-energy insertions.
The magnetic mass to this order is obtained as $\approx 2.38Cg^2T/4\pi$
where $g$ is coupling constant, $T$ is the temperature and
$C$ is defined by $C\delta^{ab}=f^{amn}f^{bmn}$, $f^{amn}$ being the structure
constants of the gauge group.($C=N$ for an $SU(N)$-gauge theory.)
An immediate question is whether the two- and higher
loop contributions are smaller than the one-loop terms. Purely
based on counting dimensions of momentum integrals and powers
of $g$ , we cannot conclude whether higher loop effects are
smaller or not. It is really an issue of numerical factors
and possible logarithms of $g$. Now, the magnetic mass, by the
standard arguments of dimensional reduction at high
temperatures is related to the mass gap of the three-dimensional
QCD. For the mass gap of three-dimensional gauge theories,
the one-loop calculations can hardly be adequate. However, for the
quark-gluon plasma, perturbative loop expansion, perhaps
with resummations, is expected to be reasonable at high
temperatures, at least for a number of processes of interest
and our approach is applicable. In any case, qualitatively, it is
interesting that a nonzero value can be obtained to this order;
computationally, it is useful to have a loopwise implementation
of gauge-invariant magnetic screening, irrespective of the specific
numerical value.

In the rest frame of the plasma, the magnetic mass term considered in [6]
has the following form
$$\eqalignno{
{\tilde \Gamma}&=-M^2S_m&(1a)\cr
S_m&=\int d\Omega ~K(A_n,A_{\bar n})&(1b)\cr}
$$
where $A_n={1\over 2}A_in_i, A_{\bar n}={1\over 2} A_i{\bar n}_i$.
$n_i$ is a (complex) three-dimensional null vector of the form
$$n_i=(-\cos\theta \cos\vf-i\sin\vf,-\cos\theta
\sin\vf+i\cos\vf,\sin\theta)\eqno(2)$$
In Eq.(1), $d\Omega=\sin\theta ~d\theta d\vf$ and denotes integration over
the angles of $n_i$. $K(A_n,A_{\bar n})$ is given by
$$K(A_n,A_{\bar n})=-{1\over\pi}\int d^2 x^T\biggl[\int d^2 z
{}~{\rm Tr}(A_n,A_{\bar
n})+i\pi I(A_n) + i\pi I(A_{\bar n})\biggr]\eqno(3)$$
$z=n\cdot{\vec x}, {\bar z}={\bar n}\cdot{\vec x}$,
$x^T$ denotes coordinates
transverse to $n$, i.e., ${\vec x}^T\cdot{\vec n}=0.$
Also
$$I(A_n)=i\sum_2^\infty{(-1)^m\over m}\int {d^2 z_1 \over \pi} \ldots
{d^2z_m\over\pi}{{\rm Tr}(A_n(x_1) \ldots A_n(x_m))\over {\bar z_{12}\bar
z_{23} \dots\bar z_{m-1m}\bar z_{m1}}}\eqno(4)$$
$\bar z_{ij}=\bar z_i-\bar z_j$.
The argument of all $A$'s in Eq.(4) is the same for the transverse coordinates
$x^T$. The lowest order term in $S_m$ was shown to be
$$
S_m=~\int {d^4k \over (2\pi )^4} ~\12
 A_i^a(-k) A_j^a(k)\biggl(\delta_{ij}-{k_ik_j\over
{\vec k}^2}\biggr)  +{\cal O}(A^3)\eqno(5)
$$
This term involves only the transverse potentials, as expected for
magnetic screening and on account of gauge invariance. The terms with
higher powers of $A$ make $S_m$ invariant under the full non-Abelian gauge
transformations.

The general strategy for the inclusion of this term is as
follows. We write the action as
$$
S=S_0+M^2S_m-\Delta ~S_m\eqno(6)
$$
$S_0$ is the standard quark and gluon part of the action. Below we shall
not consider the quark terms since their effects
are small and have the same general qualitative features. $S_0$ will be
just the Yang-Mills action. $\Delta$ is taken to have a loop expansion,
$\Delta=\Delta^{(1)}+\Delta^{(2)}+\ldots$. Calculations can be done in a
loop expansion. We require the pole of the propagator to remain at
$k_0^2-{\vec k}^2=M^2$ (for the transverse potentials)
as loop corrections are added. This requires choosing $\Delta^{(1)}$
to cancel the one-loop shift of the pole, $\Delta^{(2)}$ to cancel
the two-loop shift of the pole, etc., as is
usually done for mass renormalization. Of course, we do not want
to change the theory, only rearrange and resum various terms. Thus
we should impose the condition
$$
\Delta=\Delta^{(1)}+\Delta^{(2)}+\ldots=M^2\eqno(7)
$$
This condition is the gap equation determining $M$ in terms of
$g^2$ and $T$. This procedure of adding and subtracting a mass term,
with a gap equation required for consistency, is very standard,
for example, in the Nambu-Jona Lasinio model. It amounts to
the self-consistent summation of self-energy corrections.
The difference in the present case is that, for reasons of
gauge invariance, the mass term involves an infinite number
of interaction vertices as well.

We shall now turn to the explicit one-loop calculations.
For the plasma at high temperatures, the effects of magnetic
screening will be insignificant for high momentum processes.
The regime of interest involves momenta small compared to T.
The thermal part of the gluon propagator simplifies as
$$
{\delta(k^2-M^2)\over e^{\omega_k/T}-1} \approx {T\over
2{\omega_k}^2}\bigl[\delta(k_0-\omega_k)+\delta(k_0+\omega_k)\bigr]
\eqno(8)
$$
$\omega_k=\sqrt{{\vec k}^2+M^2}$. This is equivalent to using
Euclidean three-dimensional propagators, with a coupling constant
$e=\sqrt{g^2T}$; the thermal part of the loop contributions can
be done in a three-dimensional theory. This is, of course, the
standard dimensional reduction argument. The electrostatic field,
with a Debye mass of order $gT$, will also be neglected for low
momentum calculations. The relevant momenta for which this approximation
is valid will be of order $g^2T$; the coupling constant is also
self-consistently evaluated at a scale of order $g^2T$. This is all in
keeping with the assumed hierarchy of $\Lambda_{QCD}<<g^2T<<gT<<T$
for the hot quark-gluon plasma.

The action for momenta small compared to $T$ and $gT$ can be written as
$$
S=\int d^3x~\left[{1\over 4e^2}F^a_{\mu\nu}F^a_{\mu\nu}+
{M^2\over e^2}{\cal L}_m\right] - {\Delta\over e^2}\int d^3x~{\cal L}_m
\eqno(9)
$$
$F^a_{\mu\nu}=\partial_\mu A^a_\nu - \partial_\nu A^a_\mu
+f^{abc}A^b_\mu A^c_\nu$. The integral of ${\cal L}_m$ is
the three-dimensional
Euclidean version of Eqs.(1b,3,4). There is now only one transverse
coordinate $x^T$. A convenient gauge-fixing term is ${1\over 2}
\partial\cdot A(1-M^2{1\over{\partial^2}})\partial\cdot A$.
This gives the gluon propagator as
$\delta^{ab}\delta_{ij}(k^2+M^2)^{-1}$.
The sum of all one-loop contributions to the gluon polarization is
finite.
The functional integral is thus
$$
Z=\int[dA]~\det(-\partial\cdot D)
e^{-S}
\eqno(10)
$$
$D^{ab}_\mu=\partial_\mu\delta^{ab}+f^{acb}A^c_\mu$.
The action simplifies as

$$\eqalign{
S=\int ~\12 A^a_i(-\dlmb+M^2)A^a_i &+\int ~A^a_iA^b_jA^c_kf^{abc}
(v_{ijk}+V_{ijk})\cr
&+{\textstyle{1\over 4}}f^{amn}f^{abc}\int
{}~A^m_iA^n_jA^b_iA^c_j+\ldots \cr}\eqno(11)
$$
$$\eqalignno{
v_{ijk}(k,q,-(k+q) )&={i\over 6}~[(2k+q)_j\delta_{ik}-(2q+k)_i\delta_{jk}+
(q-k)_k\delta_{ij}]&(12a)\cr
V_{ijk}(k,q,-(k+q) )&= ~-i ~{M^2 \over 24\pi }\int d\Omega ~{n_in_jn_k\over
k\cdot n}\biggl[
{q\cdot\bar n\over q\cdot n}-
{(q+k)\cdot \bar n\over (q+k)\cdot n}\biggr]
&(12b)\cr}
$$
This, after angular integration, becomes
$$\eqalign{
V_{ijk}(k,q,-(k+q) )= ~-i ~{M^2\over 6}
\biggl[&{1\over k^2q^2-(q\cdot k)^2}\biggr]
\biggl[\biggl\{{q\cdot k\over k^2}-
{q\cdot (q+k)\over (q+k)^2}\biggr\}k_ik_jk_k \cr
&+{k\cdot(q+k)\over (q+k)^2}(q_iq_jk_k+ q_kq_ik_j+q_jq_kk_i)
-(q\leftrightarrow k)\biggr]\cr}\eqno(12c)
$$
Expression (12c) is the contribution from $S_m$.
The four-point vertex involves, in addition to the standard Yang-Mills
vertex of Eq.(11), a term from $S_m$ of the form
$\int d\Omega~(n_in_jn_kn_l)$.
Since $n_i$ is a null vector we get zero from this term to one loop
order. This is one of the advantages of our form of $S_m$;
because $n_i$ is null, Wick contractions at the same point, with
a propagator of the form $\delta_{ij}(k^2+M^2)^{-1}$, give zero.
Vertices of higher than quartic order do not contribute at one-loop
level. The relevant Feynman diagrams are shown in fig. 1.
The calculations are straightforward although somewhat
tedious. Despite the algebraic complexity of $V_{ijk}$, there are
many cancellations and for the one loop part of the effective action
we find
$$
\Gamma^{(1-loop)}=~ \int {d^3k \over (2\pi )^3}~ \12 A^a_i(-k)\Pi_{ij}(k)
A^a_j(k)
+\ldots\eqno(13)
$$
$$
\Pi_{ij}=\Pi_{ij}^{(1)}+\Pi^{(2)}_{ij}-
{\Delta^{(1)}\over e^2}\biggl(\delta_{ij}-{k_ik_j\over k^2}
\biggr)\eqno(14a)
$$
$$
\eqalign{\Pi_{ij}^{(1)}=C\delta_{ij}\biggl [-{M\over 2\pi}+
\int_0^1 d\alpha~\biggl\{{3\over 4\pi}\sqrt{k^2(\alpha-\alpha^2)+M^2}
-{k\over 8\pi}\sqrt{(\alpha-\alpha^2)}&\cr
-{k^2\over 16\pi}{5-2\alpha+2\alpha^2 \over \sqrt{
k^2(\alpha-\alpha^2)+M^2}}\biggr\}\biggr]&\cr
+C{k_ik_j\over 16\pi}\int_0^1 d\alpha~\biggl[{3+6(\alpha-\alpha^2)\over\sqrt{
k^2(\alpha-\alpha^2)+M^2}}-{2\over k}\sqrt{(\alpha-\alpha^2)}
\biggr]&\cr}
\eqno(14b)
$$
$$
\eqalign{\Pi^{(2)}_{ij}={C\over 4\pi}\biggl(\delta_{ij}-3{k_ik_j\over
k^2}\biggr)\int_0^1 d\alpha \biggl[ \rtm - \rtn \biggr]&\cr
-{C\over 8\pi}\biggl(\delta_{ij}+{k_ik_j\over k^2}\biggr)\int_0^1 d\alpha
\biggl [ 2\rtn-\rtm &\cr
-k\sqrt{\alpha-\alpha^2}\biggr]&\cr}\eqno(14c)$$
$\Pi^{(1)}_{ij}$
is the  contribution of the standard Yang-Mills
diagrams. $\Pi^{(2)}_{ij}$ involves the new vertex (12c), by itself and
mixed with (12a). The $\alpha$-integrals can actually be evaluated in terms of
elementary functions but we do not need the explicit form
in what follows.
The total contribution to $\Pi_{ij}$ can be written as
$$
 \Pi_{ij}=\biggl(\delta_{ij}-{k_ik_j\over k^2}\biggr)
{M^2\over e^2}
\biggl [B+\Pi(K)-{\Delta^{(1)}\over M^2}\biggr]\eqno(15a)
$$
$$
\eqalign{\Pi(K)={Ce^2\over 4\pi M}\int_0^1 d\alpha \biggl [
{9\over 2}\sqrt{K(\af-\af^2)+1-\af+\af^2}+1-2\sqrt{K(\af-\af^2)+\af^2}&\cr
-{(K-1)\over 4}{(5-2\af+2\af^2)\over\sqrt{K(\af-\af^2)+1-\af+\af^2}}
-5\sqrt{1-\af+\af^2}&\cr
-{3\over 4}{1\over\sqrt{1-\af+\af^2}}\biggr]&\cr}\eqno(15a)
$$
$$
\eqalign{B &=\biggl({Ce^2\over 4\pi M}\biggr ) \int_0^1 d\af
\biggl[5\sqrt{1-\af+\af^2}+{3\over 4}{1\over\sqrt{1-\af+\af^2}}-3
\biggr]\cr
&\approx {Ce^2\over 4\pi M}(2.384)}\eqno(15b)
$$
$K=(k^2+M^2)/M^2$.
Notice that $\Pi_{ij}$ is transverse as expected
on grounds of gauge invariance. $\Pi(K)$ vanishes at $K=0$
and has no other zeros for positive K as can be checked
graphically. (For negative $K$, $\Pi(K)$ is complex and our
calculation which uses time-ordered products is not
applicable; see ref. [4].) The one-loop corrected inverse
propagator has the form ${M^2\over e^2}\bigl[K+\Pi(K)+\bigl(B-
{\Delta^{(1)}\over M^2}\bigr)\bigr].$
Since  $\Pi(K)=0$ at $K=0$, we see that the pole of the
propagator will not be shifted if we choose $B={\Delta^{(1)}\over
M^2}$ or $\Delta^{(1)}\approx{Ce^2M\over 4\pi}(2.384)$.
The gap equation (7) then gives
$$
M\approx(2.384){Ce^2\over 4\pi}
\approx(2.384){Cg^2T\over 4\pi} \eqno(16)
$$
With this choice of $\Delta^{(1)}$,
 the  inverse propagator has the form ${M^2\over e^2}\bigl[K+\Pi(K)
\bigr]$.
The correction $\Pi(K)$ is significant compared to $K$, for small $K$;
for large $K$, it approaches the result for massless Yang-Mills theory;
this is a necessary check for any infrared cutoff.

Eventhough $\Pi_{ij}$ is transverse in agreement with gauge invariance,
it depends on the gauge fixing used for the gluon propagator in the
loop. The position of the pole and hence the gap equation do not depend
on this. One can explicitly check this. The simplest way is as follows.
In the effective action, one can have a gauge-dependent $\Pi_{ij}$;
Generally, the higher point functions are gauge dependent as well.
Physical results, such as scattering amplitudes,
are independent of the gauge fixing used. Alternatively, one can define a new
two-point function, a new three-point function, etc., which are independent
of the
gauge-fixing
and which lead to the same physical results. One shifts some of the
contribution (to the scattering process)
from the three-point vertex to the propagator; similar shifts
are done for the higher point functions as well. The amount of shift is
determined by requiring the physical results to be the same and leads to
the pinching procedure [8]. For our gauge choice,
the pinching terms arise from the diagrams shown in fig.2.
(A quark scattering process suffices to identify these terms.)

\epsfxsize=400pt\epsfbox{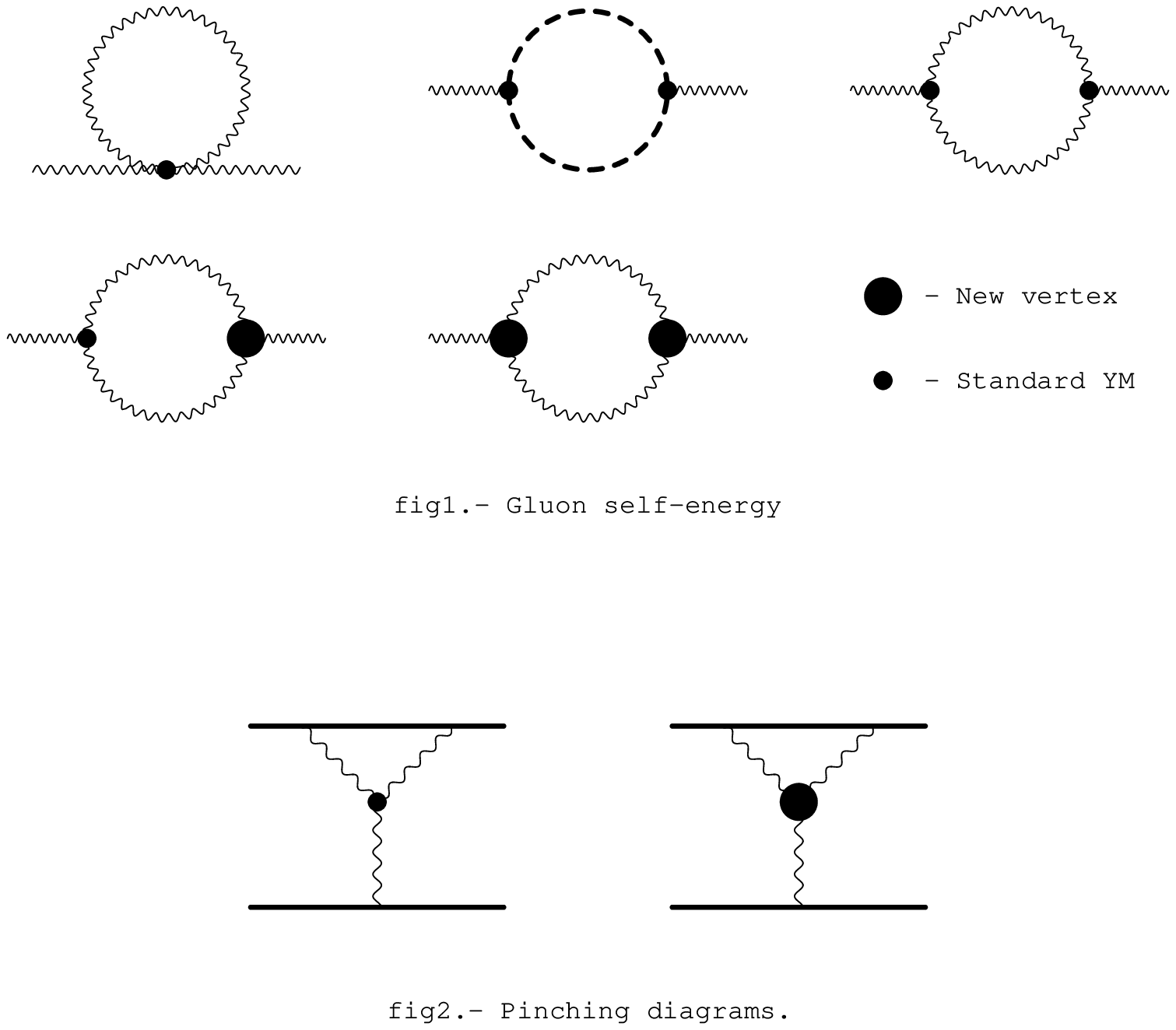}

The total pinching contribution to the two-point function is
$$
{\tilde \Pi}(K) ~= {Ce^2m \over 16 \pi }K \int_0^1 d\alpha \left[ {5\over{
\sqrt{K(\alpha -\alpha^2)+1-\alpha +\alpha^2}}}~-{2\alpha \over {\sqrt{
K(\alpha -\alpha^2) +\alpha^2}}}\right] \eqno(17)
$$
This is to be added to $\Pi (K)$ to obtain the gauge-independent
two-point vertex function. From the explicit factor of $K$ in Eq.(17),
we see that the pole remains at $K=0$.

We shall now briefly consider two-loop corrections.
These must involve $e^4$ and since $e^2$ has the dimension
of mass, the two-loop contribution to $B$ has the form
$$
B^{(2-loop)}\approx{(Ce^2)^2\over M^2}\bigl[\gamma_1+\gamma_2\log
(T/M)\bigr]\eqno(18)
$$
where $\gamma_1,\gamma_2$ are pure numbers. The gap equation now reads
$$
x^2={\textstyle{2.384\over 4\pi}}~x+ \gamma_1+
\gamma_2\log
(T/M) \eqno(19)
$$
$M=Ce^2 x$. The numerical values of $\gamma_1,\gamma_2$
determine whether the two-loop corrections are small. There are seventy-nine
diagrams for $\Pi_{ij}$ at two-loop level as opposed to five at the
one-loop level. Since the vertices given by $S_m$ are fairly
complicated we have not completed the two-loop calculation.
A preliminary analysis of some of the diagrams suggests that the two-loop
effects may be smaller by a factor of three or four. It should be noted
that this issue does not impinge on the use of $S_m$ as a
gauge-invariant infrared cutoff using the above procedure.
It only affects the numerical determination of $M$ in a
loop expansion.

\vskip .2in
\noindent{\bf References}
\vskip .2in
\item{[1]} R. Efraty and V.P. Nair, {\it Phys.Rev.Lett.} {\bf 68} (1992) 2891;
{\it Phys.Rev.} {\bf D 47} (1993) 5601.
\vskip .1in
\item{[2]} R. Pisarski, {\it Physica} {\bf A 158}(1989) 246;
{\it Phys.Rev.Lett.}
{\bf 63}, (1989) 1129; E. Braaten and R. Pisarski, {\it Phys.Rev.}
{\bf D 42} (1990) 2156; {\it Nucl.Phys.} {\bf B 337} (1990) 569;
{\it ibid.} {\bf B 339},
(1990) 310; {\it Phys.Rev.} {\bf D 45} (1992) 1827.
\vskip .1in
\item{[3]} J. Frenkel and J.C. Taylor, {\it Nucl.Phys.} {\bf B 334} (1990) 199;
J.C. Taylor and S.M.H. Wong, {\it Nucl.Phys.} {\bf B 346} (1990) 115.
\vskip .1in
\item{[4]} R. Jackiw and V.P. Nair, {\it Phys.Rev.} {\bf D 48} (1993) 4991.
\vskip .1in
\item{[5]} J.P. Blaizot and E. Iancu, {\it Phys.Rev.Lett.} {\bf 70}
(1993) 3376; {\it Nucl.Phys.} {\bf B 417} (1994) 608.
\vskip .1in
\item{[6]} V.P. Nair, Preprint CCNY-HEP 4/94.
\vskip .1in
\item{[7]} A. Billoire, G. Lazarides and Q. Shafi, {\it Phys. Lett.}
{\bf 103 B} (1981) 450;
O.K. Kalashnikov, {\it JETP Lett.}  {\bf 39} (1984) 405;
T.S. Biro and B. Muller, {\it Nucl.Phys.} {\bf A 561} (1993) 477;
O. Philipsen, in Proceedings of the NATO workshop, Sintra, Portugal,
March 1994 (to be published); W. Buchmuller and O. Philipsen, Preprint
DESY-94-202 (November 1994).
\vskip .1in
\item{[8]} J.M. Cornwall, {\it Phys.Rev.} {\bf D 26} (1982) 1453;
J.M. Cornwall and J. Papavassiliou, {\it Phys.Rev.} {\bf D 40} (1989) 3474.

\end